\documentstyle[12pt]{article}

\newtheorem{theorem}{Theorem}
\newtheorem{lemma}{Lemma}

\begin{document}

\newcommand{\lap}{\bigtriangleup}
\def\be{\begin{equation}}
\def\ee{\end{equation}}
\def\bea{\begin{eqnarray}}
\def\eea{\end{eqnarray}}
\def\beas{\begin{eqnarray*}}
\def\eeas{\end{eqnarray*}}
 
\def\R{{\rm I\kern-.1567em R}}
\def\N{{\rm I\kern-.1567em N}}
 
\def\supp{\mbox{\rm supp}\,}
\def\dist{\mbox{\rm dist}\,}

\def\ekin{E_{\rm kin}}
\def\epot{E_{\rm pot}}

\def\C{{\cal C}}
\def\X{{\cal X}}
\def\F{{\cal F}}
\def\M{{\cal M}}
\def\H{{\cal H}}

\def\prfe{\hspace*{\fill} $\Box$

\smallskip \noindent}

\title{Isotropic steady states in galactic dynamics revised}
\author{ Yan Guo\\
         Lefschetz Center for Dynamical Systems \\
         Division of Applied Mathematics \\
         Brown University, Providence, RI 02912 \\
         and \\
         Gerhard Rein\\
         Mathematisches Institut 
         der Universit\"at M\"unchen\\
         Theresienstr. 39\\
         80333 M\"unchen, Germany}
\date{}
\maketitle

\begin{abstract}
The present paper completes our earlier results
on nonlinear stability of stationary solutions of the 
Vlasov-Poisson system in the stellar dynamics case.
By minimizing the energy under a mass-Casimir constraint
we construct a large class of isotropic, spherically symmetric
steady states and prove their 
nonlinear stability against {\em general}, i.~e.,
not necessarily symmetric perturbations. 
The class is optimal in a certain sense, in particular, it includes all
polytropes of finite mass with decreasing dependence on the particle 
energy.
\footnote{This work was published as {\sc Guo, Y., Rein, G.:}
Isotropic steady states in galactic dynamics.
{\em Commun.\ Math.\ Phys.}\ {\bf 219}, 607--629 (2001). 
We correct an error in the analysis of the limiting ``Plummer case''
which was pointed out to us by Y.-H.~Wan:
In the proof of the former Lemma~7 we used the 
``identity'' $\int f^\ast (x,v) dv
= (\int f(x,v) dv)^\ast$ where $\ast$ denotes the symmetric decreasing
rearrangement with respect to $x$. This is clearly false, and we
modify Section~6 accordingly and also a detail in the proof of Theorem~3.}  
\end{abstract}

\section{Introduction}
\setcounter{equation}{0}

The question of which galaxies or globular clusters
are stable has for many years attracted
considerable attention in the astrophysics literature, 
cf.\ \cite{BT,FP} and the references there.
If one neglects relativistic effects and collisions among the
stars, then from a mathematics point of view 
the question is which steady states of the
Vlasov-Poisson system 
\[
\partial_t f + v \cdot \partial_x f - \partial_x U \cdot 
\partial_v f = 0 ,
\]
\[ 
\lap U = 4 \pi\, \rho,\ \lim_{|x| \to \infty} U(t,x) = 0 , 
\]
\[
\rho(t,x)= \int f(t,x,v)dv ,
\]
are stable.
Here $f = f(t,x,v)\geq 0$ denotes the density of the stars 
in phase space, $t \in \R$ denotes time, $x, v \in \R^3$ denote 
position and velocity respectively, $\rho$ is the spatial mass 
density of the stars, and $U$ the gravitational potential
which the ensemble induces collectively. 

If $U_0$ is a time-independent potential then 
the particle energy    
\be \label{parten}
E=\frac{1}{2}|v|^2 + U_0 (x), 
\ee
is conserved along the characteristics of the Vlasov equation. 
Therefore, a standard technique to
obtain steady states of the Vlasov-Poisson system
is to  prescribe the particle
distribution $f_0$ as a function of the particle
energy---this takes care of the Vlasov equation---and 
to solve self-consistently the
remaining Poisson equation. The main problem then is to show that 
the resulting steady state has finite mass and possibly compact support.
A well known class of steady states for which this approach works are the
so-called polytropes
\be \label{poly}
f_0 (x,v) = (E_0 - E)_+^k.
\ee
Here $(\cdot)_+$ denotes the positive part, $E_0\in \R$ is a cut-off 
energy, and  $-1/2 < k\leq 7/2$; only for this range of exponents
do these steady states have finite mass, if $k<7/2$ they have
compact support in addition. If $f_0$ depends only on the particle 
energy the resulting steady state is isotropic and spherically symmetric.
Assuming spherical symmetry of $U_0$ to begin with  
steady states may also depend on a further conserved quantity,
the modulus of angular momentum squared,
\be \label{angmom}
L:= |x|^2 |v|^2 - (x \cdot v)^2 ,
\ee
in which case they are no longer isotropic. According to Jeans' Theorem
the distribution function of any spherically symmetric steady state
has to be a function of the invariants $E$ and $L$, cf.\ \cite{BFH}. 

In \cite{G1,GR1,GR2,R3} we addressed the stability of steady states
by a variational technique: It was shown that an appropriately chosen
energy-Casimir functional has a minimizer
under the constraint that the mass is prescribed, 
this minimizer was shown to be a steady state,
and its nonlinear stability was derived from its minimizing
property. While this turned out to be an efficient method to assess
the stability of known steady states and also to construct new
ones which automatically have finite mass, compact support, and are
stable, there were two unwanted restrictions: 
Perturbations had to be spherically symmetric,
and only the polytropes with $0<k<3/2$ were covered.
A physically realistic perturbation, say by the gravitational
pull of some distant galaxy, is hardly spherically symmetric.
Also, while the restriction $k>0$ was indispensable---for
$k\leq 0$ no corresponding Casimir functional can be defined---and
is probably necessary for stability since it makes $f_0$ a decreasing 
function of the particle energy, the restriction $k<3/2$  
is less well motivated.
In \cite{G2} the first author removed the latter restriction
in the case of the polytropes, 
while in \cite{R4} the second author removed the restriction
to spherically symmetric perturbations for a class of isotropic steady
states including the polytropes with $k< 3/2$ .
It is the purpose of the present paper to combine these techniques
to obtain a result which we believe is optimal in the following sense:
It does not require any symmetry restrictions of the perturbations,
and it covers all isotropic polytropes. As a matter
of fact, the restrictions we require of the steady states are
necessary to guarantee finite mass and to make the distribution
function a decreasing function of the energy. Presumably, if the
latter condition is violated sufficiently strongly, then the steady
state is unstable.

The new elements in our analysis which allow for the improvements
described above are the following: Previously we minimized an
energy-Casimir functional under a mass constraint. Now we minimize
the total energy of the system under a mass-Casimir constraint.
The change of the role of the Casimir functional---from
part of the minimized functional into part of the constraint---allows
us to remove the restriction $k<3/2$ and was introduced in \cite{G2}.
It also leads to a much cleaner assumption on the steady state
or on the Casimir functional respectively.
Inspired by the 
concentration-compactness argument due to {\sc P.~L.~Lions} \cite{L},
which was used in \cite{R4}, we use a refinement of our previous scaling 
and splitting argument in the 
compactness analysis of the energy 
functional to get rid of the symmetry assumption for the perturbations. 

The investigation is restricted to isotropic steady states: 
If one includes anisotropic ones then the Casimir functional 
is not conserved along not spherically symmetric solutions, and 
the method breaks down. Anisotropic steady states are---under the
appropriate assumptions---stable against spherically symmetric 
perturbations. Whether or not they are stable against general
perturbations remains an open problem.

The paper proceeds as follows: In the next section we establish 
some preliminary estimates which in particular 
show that the total energy is bounded from below and 
the  kinetic energy is bounded along minimizing sequences.
In Section~3 the existence of a minimizer of the energy is established.
To prevent mass from running off to spatial infinity along a minimizing
sequence we analyze how the total energy behaves under scaling 
transformations and under splittings of the distribution into 
different pieces.
In Section~4 we show that such minimizers are spherically 
symmetric steady states of the Vlasov-Poisson
system with finite mass and compact support.
The stability properties of the steady states are then discussed in the 
Section~5. Here we point out one problem: 
If $f_0$ is a steady state then 
$f_0(x+t\,V, v+V)$ for any given velocity $V \in \R^3$ is a
solution of the Vlasov-Poisson system which for $V$ small starts
close to $f_0$, but travels away from $f_0$ at a linear
rate in $t$. This trivial ``instability'', which cannot be present
for spherically symmetric perturbations, is handled
by comparing $f_0$ with an appropriate shift in $x$-space
of the time dependent perturbed solution $f(t)$.
Technically, the necessity of this shift arises in the application 
of our compactness argument. 
The considerations discussed so far are
restricted to Casimir functionals satisfying a growth condition which
excludes the polytropic case $k=7/2$. This limiting case, the
so-called Plummer sphere, is investigated in the final section.
It poses additional difficulties
due to a particular scaling invariance of the various
functionals considered, but by using rearrangements we are able 
to reduce it
to the same problem {\em with} symmetry, which has been investigated 
in \cite{G2}.  This shows that is can be essential to understand
the symmetric case first, and we hope that such a reduction to 
symmetry can be applied to other problems as well.

We conclude the introduction with some references, where we also compare
our approach with other approaches to the stability problem.
The first nonlinear stability result for the Vlasov-Poisson
system in the present stellar dynamics case is due to {\sc G.~Wolansky}
\cite{Wo}. It is restricted to spherically symmetric perturbations
of the polytropes  
\be \label{polyl}
f_0 (x,v) = (E_0 - E)_+^k L^l
\ee
with exponents $l > -1$, $0< k < l+3/2$ with $k \neq - l - 1/2$
and uses a variational approach for a reduced functional which is
not defined on a set of phase space densities $f$ but on a set
of mass functions $M(r):=\int_{|y| \leq r} \rho(y)\,dy$ with $r \geq 0$ 
denoting the radial
coordinate. In particular, is does not yield a stability estimate
directly for the phase space distribution $f$.
In \cite{Wa} {\sc Y.-H.~Wan} proves stability
by a careful investigation of the quadratic and higher order terms 
in a Taylor expansion of the energy-Casimir functional 
about a steady state. He has to assume the
existence of the steady state, requires a strong condition on $f_0$
which is satisfied by the polytropes only for $k=1$ and $l=0$,
but his arguments do not require spherical symmetry of the admissible 
perturbations.
We also mention \cite{A} where stability for the limiting 
polytropic case $k=7/2$ and
$l=0$ is considered.
Global classical solutions 
to the initial value problem for the Vlasov-Poisson system were first 
established in \cite{P}, cf.\ also \cite{S}.  
A rigorous  result on linearized stability 
is given in \cite{BMR}. For the plasma physics case,
where the sign in the Poisson equation is
reversed, the stability problem is better
understood; we refer to \cite{BRV,GS1,GS2,R1}.  
Finally, a very general condition
which guarantees finite mass and compact support of steady
states, but not their stability, is established in \cite{RR}.
   
\section{Preliminaries}
\setcounter{equation}{0}

For a measurable function $f=f(x,v)$ we define
\[
\rho_f (x):= \int f(x,v)\, dv,\ x \in \R^3,
\]
and
\[  
U_f :=  - \rho_f \ast \frac{1}{|\cdot|}.
\]
Next we define
\beas
\ekin (f)
&:=&
\frac{1}{2} \int\!\!\int |v|^2 f(x,v)\,dv\,dx \\
\epot (f)
&:=&
- \frac{1}{8\pi} \int |\nabla U_f (x)|^2 dx = 
- \frac{1}{2} \int\!\!\int\frac{\rho_f(x) \rho_f(y)}{|x-y|}dx\,dy ,\\
\H(f) 
&:=&
\ekin(f)  + \epot (f),
\eeas
and
\[
\C(f)
:=
\int\!\!\int Q(f(x,v))\,dv\,dx 
\]
where 
$Q$ is a given function satisfying certain assumptions specified 
below.
We will minimize the total energy or Hamiltonian $\H$ of the
system under a mass-Casimir constraint, i.~e., over the set
\be \label{spacedef}
\F_M := \Bigl\{ f \in L^1(\R^6) 
\mid
f \geq 0,\ \C(f) = M,\ \ekin(f) < \infty\Bigr\},
\ee
where $M>0$ is prescribed. The function $Q$ has to satisfy
the following  

\smallskip
\noindent {\bf Assumptions on $Q$}: 
$Q \in C^1 ([0,\infty[)$, $Q \geq 0$, $Q(0)=0$, and
\begin{itemize}
\item[(Q1)]
$Q(f) \geq C (f+f^{1+1/k}),\ f \geq 0$,  with constants
$C>0$ and $0 < k < 7/2$,
\item[(Q2)]
$Q$ is convex. 
\end{itemize}

\smallskip
\noindent
{\bf Remark.}
\begin{itemize}
\item[(a)]
In the last section we consider the limiting case $k=7/2$
for which
\[
Q(f) := f^{9/7},\ f \geq 0 .
\]
\item[(b)]
On their support
the minimizers obtained later will satisfy the relation 
\[
\lambda_0 Q'(f_0) = E
\]
with some Lagrange multiplier
$\lambda_0<0$ and $E$ as defined in (\ref{parten}). 
Thus $f_0$ is a function of the particle energy and thus a steady state
of the Vlasov-Poisson system, 
provided this identity can be inverted.
\item[(c)]
A typical example of a function $Q$ satisfying the assumptions is
\be \label{qpoly}
Q(f) = f + f^{1+1/{k}},\ f \geq 0,
\ee
with $0 < k < 7/2$ which leads to a steady state of polytropic form 
(\ref{poly}). 
More generally, if an isotropic steady state $(f_0,U_0)$ is given
with $f_0$ of the form
\[
f_0(x,v)=\phi(E)
\]
with some function $\phi$ then the above assumptions for the Casimir 
functional hold, if $\phi(E)$ vanishes for $E$ larger than
some cut-off energy $E_0$, $\phi(E) \leq C (E_0-E)^k,\ E \leq E_0$,
where $0<k<7/2$, and $\phi'(E) < 0,\ E < E_0$.
The existence of a cut-off energy is
necessary in order that the steady
state has finite mass. The growth condition is essential for
the compactness properties of $\H$; cf.\ the difficulties in the
limiting case $k=7/2$, and note also that the polytropic ansatz with
$k>7/2$ leads to steady states with infinite mass.
Finally, it
is generally believed that steady states are unstable if the
monotonicity condition on $\phi$ is violated sufficiently strongly.
In this sense one can say that the assumptions on $Q$ are optimal.
\item[(d)]
The function
\be \label{jump}
Q(f) = \left\{
\begin{array}{ccl}
f&,&0\leq f \leq 1,\\
\frac{1}{2}(f^2 + 1)&,&f >1
\end{array} \right.
\ee
also satisfies our assumptions and leads to
\be \label{jumpp}
f_0(x,v) =
\left\{
\begin{array}{ccl}
E/E_0&,& E < E_0\\
0&,& E \geq E_0
\end{array} \right.
\ee
with some $E_0 < 0$. Thus the fact that we do not require 
$Q \in C^2(]0,\infty[)$ with $Q''> 0$ allows for
examples where $f_0$ has jump discontinuities, 
and these steady states will turn out to be dynamically
stable as well.
\end{itemize} 
 
We collect some estimates
for $\rho_f$ and $U_f$ induced by an element $f \in \F_M$.
As in the rest of the paper constants
denoted by $C$ are positive, may depend on $Q$ and $M$, and 
their value may change from line to line.

\begin{lemma} \label{rhoest}
Let $n := k +3/2$ so that $1+1/n > 6/5$. 
Then for any $f \in \F_M$ the following holds:
\begin{itemize}
\item[{\rm (a)}] 
$f \in L^{1+1/k} (\R^6)$ with
\[
\int\!\!\int f^{1+1/k} dv\,dx 
+ \int\!\!\int f dv\,dx \leq C.
\]
\item[{\rm (b)}]
$\rho_f \in L^{1+1/n} (\R^3)$ with
\beas
\int \rho_f^{1+1/n} dx 
&\leq& 
C\, \left[ \int\!\!\int f^{1+1/k}\,dv\,dx \right]^{k/n}
\left[\int\!\!\int|v|^2 f \,dv\,dx \right]^{(n-k)/n}\\
&\leq& 
C \, \ekin (f)^{\frac{3}{2 n}}
\eeas
\item[{\rm (c)}]
$U_f \in L^{6}(\R^3)$ with $\nabla U_f \in L^{2}(\R^3)$, the two forms
of $\epot (f)$ stated above are equal, and
\[
\int |\nabla U_f|^2 dx \leq
C\,\|\rho_f\|_{6/5}^2
\leq C  \ekin (f)^{1/2}.
\]
\end{itemize}
The assertions in (b) and (c) remain valid in the limiting case
$k=7/2$ where $n=5$, cf.\ Remark (a) above.
\end{lemma}
\noindent
{\bf Proof.} 
Part (a) is obvious from assumption (Q1).
Splitting the $v$-integral according to 
$|v| \leq R$ and $|v| > R$ and optimizing in $R$ yields
\[
\rho_f^{1+1/n}
\leq C \, \left( \int f^{1+1/k} dv \right)^{k/n} 
\left( \int |v|^2 f dv \right)^{(n-k)/n} .
\]
Therefore, the first estimate in (b) follows from H\"older's inequality 
with indices $n/k$ and $n/(n-k)$, 
and part (a) implies the second estimate in (b). 
Since $\rho_f \in L^1 \cap L^{1+1/n} (\R^3)$ and $1+1/n > 6/5$
we find by interpolation,
\[
\int \rho_f^{6/5} \leq C \ekin (f)^{3/10}; 
\]
in the limiting case this follows directly without interpolation.
The estimates for $U_f$ follow from the generalized Young's inequality,
and the equality of the two representations for $\epot (f)$ follows
by integration by parts after regularizing $\rho_f$ if necessary. 
\prfe

As an immediate corollary of the lemma above we note  that
on $\F_M$ the total energy $\H$ is bounded from below in such a way
that $\ekin$---and thus certain norms of $f$ and $\rho_f$---remain bounded 
along minimizing sequences:

\begin{lemma} \label{lower}
There exists a constant $C>0$ such that
\[
\H (f) \geq  \ekin (f) - C \ekin (f)^{1/2},\ f \in \F_M,
\]
in particular,
\[
h_M := \inf_{\F_M} \H  > - \infty,
\]
and $\ekin$ is bounded along minimizing sequences of $\H$ in $\F_M$.
\end{lemma}

The behavior of $\H$ and $\C$ under scaling transformations can be
used to show that $h_M$ is negative and to
relate the $h_M$'s for different values of $M$:

\begin{lemma} \label{scaling}
\begin{itemize}
\item[{\rm (a)}]
Let $M>0$. Then $-\infty < h_M  < 0$.
\item[{\rm (b)}]
For all $M ,\ \overline{M} > 0$,
\[
h_{\overline{M}} = \left( \overline{M}/M\right )^{7/3} h_{M}.
\]
\end{itemize}
\end{lemma}
{\bf Proof.} 
Given any function $f$, we define a rescaled function 
$\bar f(x,v)=f(ax,bv)$, where $a,\,b >0$. Then
\be \label{casscale}
\C (\bar f) =
\int\!\!\int Q(f(ax,bv))\,dv\,dx = (a\,b)^{-3} \C (f)
\ee
i.~e.\ $f \in \F_M$ iff $\bar f \in \F_{\overline{M}}$ where
$\overline{M} := (a b)^{-3} M$. Next
\beas
\ekin (\bar f) 
&=&
\frac{1}{2} \int\!\!\int |v|^2 f(ax,bv)\,dv\,dx
= a^{-3} b^{-5}\ekin(f),  \\
\epot (\bar f) 
&=&  
- \frac{1}{2} \int\!\!\int\!\!\int\!\!\int
\frac{f(ax,bv)\, f(ay,bw)}{|x-y|} dw\,dv\,dy\,dx
= a^{-5} b^{-6} \epot (f)  .
\eeas
To prove (a) we fix any $f \in \F_M$ and
let $a=b^{-1}$ so that $\bar f \in \F_M$ as well.
Then
\[
\H (\bar f) = b^{-2} \ekin (f) + b^{-1} \epot (f) < 0
\]
for $b>0$ sufficiently large, since $\epot (f) < 0$.
To prove (b) choose $a$ and $b$ such that $a^{-3} b^{-5}=a^{-5} b^{-6}$,
i.~e., $b=a^{-2}$. Then 
\be
\H (\bar f) = a^7 \H (f), \label{hscale}
\ee
and since $a = (\overline{M}/M)^{1/3}$ and the mapping 
$\F_M\to \F_{\overline{M}}$, $f \mapsto \bar f$
is one-to-one and onto this proves (b).   
\prfe 
One should note that both Lemma~\ref{lower} and Lemma~\ref{scaling}
remain valid in the limiting case $k=7/2$.
 
\section{Existence of minimizers for $k<7/2$}
\setcounter{equation}{0}

It is conceivable that along a minimizing sequence the mass
could run off to spatial infinity and/or spread uniformly in space.
The main problem in proving the existence of a minimizer is 
to show that this does not happen, which is 
done in the next lemma. 
Combined with a local compactness result for the
induced fields and a new version of the splitting technique
developed in our previous papers this will yield the existence of 
minimizers. 

\begin{lemma} 
\label{concentrate}
Let $(f_i) \subset \F_M$ be a minimizing sequence of $\H$.
Then there exist a sequence
$ (a_i) \subset \R^3$ and $\epsilon_0>0,\ R_0>0$ such that 
\[
\int_{a_i+B_{R_0}} \int Q(f_i)\, dv\,dx \geq \epsilon_0
\]
for all sufficiently large $i \in \N$.
Here we define 
$B_R := \{ x\in \R^3||x|\le R\}$.
\end{lemma}

\noindent
{\bf Proof.} For $R>1$ define
\[
K_R(x):= \left\{ \begin{array}{ccl}
R &,& |x| < 1/R,\\
1/|x|&,& 1/R \leq |x| \leq R,\\
0 &,& |x| > R ,
\end{array} \right.
\]
and
\[
F_R (x) := \frac{1}{|x|} {\bf 1}_{\{|x| > R\}} (x),\ 
G_R (x) := \left(\frac{1}{|x|} - R \right) {\bf 1}_{\{|x| < 1/R\}} (x)
\]
so that we split the kernel
\be \label{splitgreen}
\frac{1}{|x|} = K_R (x) + F_R (x) + G_R (x),\ x \in \R^3 .
\ee
Here ${\bf 1}_A$ denotes the indicator function of the set $A$.
We split
\be \label{i}
\frac{1}{4 \pi} \int |\nabla U_i|^2 dx 
= \int\!\!\int \frac{\rho_i (x) \rho_i (y)}{|x-y|} dy\,dx = I_1 + I_2 + I_3
\ee
according to (\ref{splitgreen}), where $\rho_i := \rho_{f_i}$. 
Since $(\rho_i)$ is bounded in $L^{1+1/n}(\R^3)$ and by (Q1) also in
$L^1 (\R^3)$, we find from Lemma~\ref{rhoest} (b), 
using the boundedness of the
kinetic energy,
\begin{eqnarray}
|I_1| 
&\leq&
R \,
\int\!\!\int_{|x-y| < R} \rho_i(x)\,\rho_i(y)\,dx\,dy 
\leq R C \sup_{y \in \R^3} \int_{y+B_R} \rho_i (x)\, dx \label{i1}\\
&\le& R ^{(n+4)/(n+1)}C \,\sup_{y \in \R^3}
\left[ \int_{y+B_R} \rho_i ^{1+1/n} dx\right]^{n/(n+1)} \nonumber\\
&\le&
R ^{(n+4)/(n+1)} C\, 
\sup_{y \in \R^3} 
\left[\int_{y+B_R} \int f_i^{1+1/k}\,dv\,dx \right]^{k/(n+1)},\nonumber 
\end{eqnarray}
and
\beas 
|I_2|
&\leq&
\frac{1}{R} \int\!\!\int \rho_i(x)\,\rho_i(y)\,dx\,dy \leq C \, R^{-1},
\\
|I_3|
&\leq&
\|\rho_i\|_{1+1/n} \|\rho_i \ast G_R\|_{n+1} \leq 
C \,\|G_R\|_{(n+1)/2} \leq
C \, R^{-(5-n)/(n+1)};
\eeas
for the last estimate we used H\"older's and Young's inequality. 
Since $(f_i)$ is a minimizing sequence we have, for any $R>1$,   
\be
h_M/2>{\cal H}(f_i)\ge -| I_1|-|I_2|-|I_3|, \label{1/2}
\ee
provided $i$ is sufficiently large.
Therefore,
\bea
&&
\liminf_{i\to \infty}
\left[\sup_{y\in \R^3}\int_{y+B_R}
\int f_i^{1+1/k}\,dv\,dx \right]^{k/(n+1)} =
\liminf_{i\to \infty} \,|I_1| \, R^{-(n+4)/(n+1)}\nonumber \\
&&
\hspace*{3.5cm}
\geq
C\, R^{-(n+4)/(n+1)}
\left[ - h_M /2-R^{-1} - R^{-(5-n)/(n+1)} \right].\label{ii} 
\eea 
By Lemma~\ref{scaling} (a) the right hand side of this estimate
is positive for $R$ sufficiently large, and the proof is complete.
\prfe

\begin{lemma} \label{compact}
Let $(\rho_i) \subset L^1 \cap L^{1+1/n} (\R^3)$ be bounded with respect 
to both norms and $\rho_0 \in L^1 \cap L^{1+1/n} (\R^3)$ with
\[
\rho_i \rightharpoonup \rho_0 \ \mbox{weakly in}\ L^{1+1/n} (\R^3) .
\]
Then for any $R>0$,  
\[
\nabla U_{{\bf 1}_{B_R} \rho_i} \to \nabla U_{{\bf 1}_{B_R} \rho_0} 
\ \mbox{strongly in}\ L^2 (\R^3).
\]
\end{lemma}

\noindent
{\bf Proof}. Take any $R'>R$. 
Since by assumption on $k$ we have $1+1/n \in ]6/5,5/3[$,
the mapping
\[
L^{1+1/n} (\R^3) \ni \rho \mapsto {\bf 1}_{B_{R'}} \nabla U_{\rho} 
\in L^2(B_{R'})
\]
is compact. Thus the asserted strong convergence holds on $B_{R'}$.
On the other hand,  
\[
\int_{|x| \geq R'} |\nabla U_{{\bf 1}_{B_R} \rho_i}|^2 dx
\leq \frac{C}{R'-R} \|\rho_i\|_1^2 \leq \frac{C}{R'-R},
\ i \in \N \cup \{0\},
\]
which is arbitrarily small for $R'$ large.\prfe

We are now ready to show the existence of a minimizer of $\H$.

\begin{theorem} \label{exminim}
Let $M>0$.
Let $(f_i) \subset \F_M$ be a minimizing sequence of 
$\H$. Then there is a minimizer $f_0\in \F_M$, a subsequence (still 
denoted by 
$(f_{i})$), and a sequence of translations 
$T_i f_{i}(x,v)=f_i(x+a_i,v)$ 
with $(a_i) \subset \R^3$, such that 
\[
\H (f_0) = \inf_{\F_M} \H = h_M
\] 
and $T_i f_i \rightharpoonup f_0$ weakly in 
$L^{1+1/k} (\R^6)$.
For the induced potentials we have
$\nabla U_{T_i f_i} \to \nabla U_0$ strongly in $L^2 (\R^3)$.
\end{theorem}

\noindent
{\bf Remark.} Without admitting shifts in $x$-space the assertion
of the theorem is wrong: Starting from a given minimizer $f_0$
and a sequence of shift vectors $(a_i) \in \R^3$ the sequence
$(T_i f_0)$ is minimizing and in $\F_M$, but if $|a_i| \to \infty$
this minimizing sequence converges  weakly to zero, which is not in $\F_M$. 

\noindent
{\bf Proof of Theorem~\ref{exminim}.} 
Let $(f_i)$ be a minimizing sequence and
$(a_i) \subset \R^3$ such that the assertion of Lemma~\ref{concentrate}
holds. 
Since $\H$ is translation invariant
$(T_i f_i)$ is again a minimizing sequence.
By Lemma~\ref{rhoest} (a), $(T_i f_i)$ is bounded in $L^{1+1/k} (\R^6)$. 
Thus there exists a weakly convergent
subsequence, denoted by $(T_i f_i)$ again:
\[
T_i f_i \rightharpoonup f_0\ \mbox{weakly in }\ L^{1+1/k} (\R^6).
\]
Clearly, $f_0 \geq 0$ a.~e.
By Lemma~\ref{lower}, $(\ekin( T_i f_i))$ is bounded  so
by Lemma~\ref{rhoest}, $(\rho_i)=(\rho_{T_i f_i})$ is bounded in 
$L^{1+1/n} (\R^3)$, and by assumption (Q1) this sequence is also 
bounded in $L^1 (\R^3)$.
After extracting a further subsequence
\[
\rho_i \rightharpoonup \rho_0:=\rho_{f_0} \
\mbox{weakly in }\ L^{1+1/n} (\R^3) .
\]
Also by weak convergence 
\[
\ekin(f_0) \leq \liminf_{i \to \infty} \ekin(T_i f_i) < \infty.
\] 
By (Q2) the functional $\C$ is convex.
Thus by Mazur's Lemma and Fatou's Lemma  
\[
\C(f_0) \leq \limsup_{i \to \infty}\, \C(T_i f_i) = M, 
\]
in particular, $\rho_0 \in L^1(\R^3)$ by (Q1). 

The key step is to show  that up to a subsequence we have
\be \label{strong}
\|\nabla U_{T_i f_i}-\nabla U_0 \|_2\to 0.
\ee
For $R_0<R$ we denote $B_{R_0,R}:=\{x \in \R^3 | R_0 \leq |x| \leq R \}$, 
and we split $T_if_i$ as follows:
\bea
T_if_i&=&T_i f_i{\bf 1}_{B_{R_0} \times \R^3} +
T_i f_i{\bf 1}_{B_{R_0,R}\times \R^3}
+T_i f_i{\bf 1}_{B_{R,\infty} \times \R^3} \nonumber\\
&=:& f_i^1+f_i^2+f_i^3 \label{fsplit}.
\eea
Due to Lemma \ref{compact}, $\nabla U_{f_i^1+f_i^2}$ converges strongly 
in $L^2$ for any fixed $R$. It thus suffices to show that for any 
$\epsilon>0$, 
\be \label{leak1}
\liminf_{i\to\infty} \int |\nabla U_{f_i^3}|^2 dx < \epsilon
\ee
for sufficiently large $R$. By Lemma~\ref{rhoest} (b) 
we only need to show that 
\be \label{leak2}
\liminf_{i\to\infty} \int\!\!\int Q(f_i^3) \,dv\,dx < \epsilon
\ee
for sufficiently large $R$.  
We use the method of splitting to verify (\ref{leak2}). 
According to (\ref{fsplit}),
\bea
\H (T_if_i)
&=&
\H (f_i^1) + \H (f_i^2) + \H (f_i^3)
\nonumber \\
&&
{}
-\int\!\!\int {{\rho_i^2(x) (\rho_i^1+\rho_i^3)(y)}\over{|x-y|}}dx\,dy-
\int\!\!\int {{\rho_i^1(x)\rho_i^3(y)}\over{|x-y|}}dx\,dy \nonumber
\\
&=:& 
\H (f_i^1) + \H (f_i^2) + \H (f_i^3)- I_1 - I_2, \label{split}
\eea
with obvious definitions for $\rho_i^1, \rho_i^2, \rho_i^3$.
The boundedness of $\|\nabla U_{\rho_i^1+\rho_i^3}\|_2$ implies that 
\[
I_1 \leq C\,\|\nabla U_{\rho_i^2}\|_2.
\] 
Since $\rho_i^2$ converges weakly in $L^{1+1/n}$
to $\rho_0^2 := \rho_0 {\bf 1}_{B_{R_0,R}}$, 
\be \label{middle}
\|\nabla U_{\rho_i^2}- \nabla U_{\rho_0^2} \|_2\to 0,\
i\to\infty
\ee 
by Lemma~\ref{compact}.
For $R>2R_0$ we use H\"older's inequality to estimate $I_2$ as follows:
\[
I_2
\leq
2\, \int_{B_{R_0}}\rho_i (x)dx
\int_{B_{R,\infty}}|y|^{-1}\rho_i(y)dy
\leq
C\,\|\rho_i \|_{6/5}^2\left({{R_0}\over R}\right)^{1/2}.
\]
It is a simple calculus exercise to show that
\[
\tau^{7/3} + (1-\tau)^{7/3} \leq 1 - \frac{7}{3} \tau ( 1-\tau),\ 
\tau \in [0,1].
\]
With Lemma~\ref{scaling} and obvious definitions
of $M_i^1, M_i^2, M_i^3$ this implies that
\beas
\H(f^1_i) + \H(f^2_i) + \H(f^3_i)
&\geq&
h_{M_i^1} + h_{M_i^2} + h_{M_i^3}\\
&=&
\left[\left(\frac{M^1_i}{M}\right)^{7/3} + \left(\frac{M^2_i}{M}\right)^{7/3}
+ \left(\frac{M^3_i}{M}\right)^{7/3}\right]\, h_M\\
&\geq&
\left[\left(\frac{M^1_i + M^2_i}{M}\right)^{7/3} + 
\left(\frac{M^3_i}{M}\right)^{7/3} \right] \, h_M \\
&\geq&
\left[1 - \frac{7}{3} \frac{M^1_i + M^2_i}{M}\frac{M^3_i}{M} \right]\,h_M
\eeas
and thus
\beas
h_M - \H (T_i f_i) - C_1 h_M M_i^1 M_i^3
&\leq&
I_i^1 + I_i^2 \\
&\leq&
C_2 \left[\|\nabla U_{f_0^2}\|_2 + 
\|\nabla U_{\rho_i^2} - \nabla U_{\rho_0^2}\|_2
+ \left(\frac{R_0}{R}\right)^{1/2} \right].
\eeas
Here $R> 2 R_0$ are so far arbitrary, 
and the constants $C_1, C_2$ are independent of $R$ 
and $R_0$.
Now assume (\ref{leak2}) were false. Then there exists $\epsilon_1 >0$
such that for every $R>0$ and $i$ large we have 
\be \label{leak}
\int\!\!\int Q(f_i^3)\,dv\,dx \ge \epsilon_1 .  
\ee
Define
\[
\epsilon_2 := -C_1 h_M \epsilon_0 \epsilon_1 > 0
\]
where $\epsilon_0$ is as in Lemma~\ref{concentrate}, and increase $R_0$
from that lemma such that 
$C_2 \|\nabla U_{f_0^2}\|_2 \leq \epsilon_2/4$. Next choose $R> 2 R_0$
such that $ C_2 (R_0/R)^{1/2} \leq \epsilon_2/4$. Then for $i$ large,
\beas
h_M - \H(T_i f_i) + \epsilon_2 
&\leq&
h_M - \H(T_i f_i) - C_1 h_M M_i^1 M_i^3 \\
&\leq&
\frac{1}{2} \epsilon_2 + C_2 \|\nabla U_{\rho_i^2} - \nabla U_{\rho_0^2}\|_2 .
\eeas
By (\ref{middle}) this contradicts the fact that $(T_i f_i)$ is minimizing.
Thus (\ref{leak2}) holds, and (\ref{strong}) follows.

Clearly we have $\H (f_0)\le \lim_i \H (T_if_i)$, and it remains
to show that $\C(f_0)=M$. 
Assume that $M_0 := \C(f_0) < M$; $M_0 > 0$ 
since otherwise $f_0 = 0$ in contradiction to $\H (f_0) < 0$. Let
\[
b:= \left(\frac{M_0}{M}\right)^{2/3} < 1,\ a:=b^{-1/2},
\]
so that by
(\ref{casscale}), $\bar f_0 \in \F_M$. Then by (\ref{hscale}),
\[
\H(\bar f_0) = a^7 \H(f_0) = b^{-7/2} h_M < h_M,
\]  
a contradiction; recall that $b<1$ and $h_M <0$.  \prfe

\section{Properties of minimizers}
\setcounter{equation}{0}

The purpose of the present section is to show that the minimizers
obtained in the previous one are indeed steady states of the 
Vlasov-Poisson system.

\begin{theorem} \label{propminim}
Let $f_0 \in \F_M$ be a minimizer of $\H$. Then
\[
f_0 (x,v)=\left\{
\begin{array}{ccl} 
\phi (E) &,& E < E_0,\\
0 &,& E \geq E_0
\end{array}
\right. \ \ \mbox{a.~e.}
\]
where 
\[
E := \frac{1}{2} |v|^2 + U_0 (x),
\]
\[
E_0 := \lambda_0 Q'(0),\ \lambda_0 := \frac{\int\!\!\int E\,f_0\,dv\,dx}
{\int\!\!\int Q'(f_0)\, f_0 \,dv\,dx} < 0, 
\]
$U_0$ is the potential induced by $f_0$, and
\[
\phi (E) := \inf \{ f \geq 0 | Q'(f) = E/\lambda_0 \},\ E \leq E_0.
\]
In particular, $f_0$ is a steady state of the Vlasov-Poisson system.
\end{theorem}

\newpage
\noindent
{\bf Remark.}
\begin{itemize}
\item[(a)]
The Euler-Lagrange equation for our constrained minimization problem
will give us the relation
\[
\lambda_0 Q'(f_0) = E\ \mbox{on}\ f_0^{-1} (]0,\infty[),
\]
which we want to invert by means of the function $\phi$.
Clearly, if $Q'$ is strictly increasing then
\[
\phi (E) = (Q')^{-1} (E/\lambda_0),\ E \leq E_0 .
\]
\item[(b)]
Under our general assumption on $Q$ the function
$Q' : [0,\infty[ \to [Q'(0),\infty[$
is continuous, increasing, and onto. This implies that for every
$\eta \geq Q'(0)$ the set $(Q')^{-1} (\eta)$ is a closed, bounded interval,
and there exists an at most countable set $V_{\rm crit}$ such that
$(Q')^{-1} (\eta)$ consists of one point for $\eta \notin V_{\rm crit}$.
The function $\phi$ is decreasing with $\phi(]-\infty,E_0]) = [0,\infty[$,
and for 
$f \in [0,\infty[$ with $\lambda_0 Q'(f) \notin V_{\rm crit}$
we have $\phi (\lambda_0 Q'(f)) = f$ as desired.
\item[(c)]
In the example given by (\ref{jump}),
\[
Q'(f) =
\left\{
\begin{array}{ccl}
1 &,& 0\leq f \leq 1,\\
f &,& f > 1  
\end{array}
\right.
\]
which is not one-to-one on $[0,\infty[$, but
the Euler-Lagrange equation can be inverted to yield (\ref{jumpp}).
\end{itemize}

\noindent
{\bf Proof of Theorem~\ref{propminim}.}
Let $f_0$ and $U_0$ be a pointwise defined representative
of a minimizer of $\H$ in $\F_M$ and its induced potential respectively;
to derive the Euler-Lagrange relation we will argue
first on $f_0^{-1}(]0,\infty[)$ and then
on the complement. 

For $\epsilon >0$ small, 
\[
K_\epsilon := \left\{ (x,v) \in \R^6 \mid \epsilon \leq f_0(x,v) \leq 
\frac{1}{\epsilon} \right\}
\]
defines a set of positive, finite measure.
Let $w\in L^\infty(\R^6)$ be compactly supported
and non-negative outside $K_\epsilon$, and
define
\[
G(\sigma,\tau) := \int\!\!\int 
Q(f_0 + \sigma {\bf 1}_{K_\epsilon} + \tau w )\, dv \, dx;
\]
for $\tau$ and $\sigma$ close to zero, $\tau \geq 0$, the function 
$f_0 + \sigma {\bf 1}_{K_\epsilon} + \tau w $ is bounded
on $K_\epsilon$, and non-negative otherwise. Therefore,
$G$ is continuously differentiable for such $\tau$ and $\sigma$, and
$G(0,0) = M$. Since 
\[
\partial_\sigma G (0,0) = \int\!\!\int_{K_\epsilon} Q'(f_0)\, dv\, dx \neq 0,
\]
there exists by the implicit function theorem a continuously differentiable
function $\tau \mapsto \sigma(\tau)$ with $\sigma(0)=0$, 
defined for $\tau\ge 0$ small, such that
$G(\sigma(\tau),\tau) = M$. Hence
$f_0 + \sigma (\tau){\bf 1}_{K_\epsilon} + \tau w \in \F_M$.
Furthermore, 
\be \label{sigmader}
\sigma' (0) = -\frac{\partial_\tau G(0,0)}{\partial_\sigma G(0,0)}
= - \frac{\int\int Q'(f_0)w}{\int\int_{K_\epsilon} Q'(f_0)}.
\ee
Since $G(\sigma(\tau),\tau)$ attains its
minimum at $\tau=0$,  
Taylor expansion implies 
\[
0 \leq
\H(f_0 + \sigma(\tau) {\bf 1}_{K_\epsilon}+ \tau  w ) - \H(f_0) 
= \tau
\int\!\!\int  E\, [\sigma'(0){\bf 1}_{K_\epsilon} +w]\, dv\,dx +o(\tau)
\]
for $\tau\ge 0$ small.
With (\ref{sigmader}) we get 
\be \label{vareq}
\int\!\!\int [-\lambda_\epsilon Q'(f_0)+E]\,w\, dv\,dx \ge 0
\ee
where
\[
\lambda_\epsilon := 
\frac{\int\!\!\int_{K_\epsilon}E}{\int\!\!\int_{K_\epsilon}Q'(f_0)}. 
\]
By our choice for $w$ this implies that 
$E=\lambda_\epsilon Q'(f_0)$ a.~e.\ on $K_\epsilon$ and 
$E\geq \lambda_\epsilon Q'(f_0)$ otherwise. 
This shows that $\lambda_\epsilon = \lambda_0$ does in fact not 
depend on $\epsilon$. Letting $\epsilon\to 0$,  we conclude that
\bea
E
&=&
\lambda_0 Q'(f_0) 
\ \mbox{a.~e.\ on}\ f_0^{-1}(]0,\infty[), \label{elsupp}\\
E 
&\geq& 
\lambda_0 Q'(0) = E_0 \ \mbox{a.~e.\ on}\ f_0^{-1}(0) \label{elcsupp}.
\eea
If we multiply (\ref{elsupp}) by $f_0$ and integrate
we obtain the asserted formula for $\lambda_0$, and  
$\lambda_0 < 0$ as claimed, since
\[
\int\!\!\int E \, f_0\, dv\, dx = \ekin(f_0) - 2 \epot(f_0) < \H(f_0) < 0 .
\]
We need to invert (\ref{elsupp}). Let $V_{\rm crit}$ be the at most
countable set of values where $Q'$ is not one-to-one, cf.~part (a) of
the remark above. Since for any constant $\eta \in \R$ the set $E^{-1}(\eta)$
has measure zero---for fixed $x$ this is a sphere in $v$-space---we conclude
that
\[
S_{\rm crit} := \{(x,v)\in \R^6 | E(x,v)/\lambda_0 \in V_{\rm crit} \}
\]
is a set of measure zero, and on $f_0^{-1}(]0,\infty[) \setminus S_{\rm crit}$
the Euler-Lagrange equation (\ref{elsupp}) can be inverted to yield
\[
f_0(x,v) = \phi(E)
\]
as claimed, cf.~part (a) of the remark above. Together with (\ref{elcsupp})
this proves that $f_0$ is a.~e.\ equal to a function of the 
particle energy $E$. 
\prfe

Next we study the regularity, symmetry, and uniqueness of minimizers. 
Let $C^m_c$ and $C^m_b$ denote the space
of $C^m$ functions with compact support and with bounded derivatives
up to order $m$, respectively. 

\begin{theorem} \label{regminim}
\begin{itemize}
\item[{\rm (a)}]
Let $f_0 \in \F_M$ be a minimizer of $\H$.
Then $f_0$
is spherically symmetric with respect to some point in $x$-space.
\item[{\rm (b)}]
If $k\geq 1/2$ assume in addition that
\[
\phi(E) \leq C (-E)^k,\ E \to -\infty,
\]
where $\phi$ is defined by $Q$ as in Theorem~\ref{propminim};
this condition 
is compatible with the general assumptions on $Q$. 
Then $U_0\in C^2_b (\R^3)$
with $\lim_{|x| \to \infty} U_0(x) =0$ and $\rho_0 \in C^1_c (\R^3)$. 
\item[{\rm (c)}]
If in particular $Q(f) = f+ f^{1+1/k},\ f \geq 0$,
with $0 < k < 7/2$ then up to a shift 
in $x$-space there are at most two minimizers of $\H$ in $\F_M$.
\end{itemize}
\end{theorem}

\noindent
{\bf Proof.}
To prove the spherical symmetry of $f_0$ we denote by $f_0^\ast$
the spherically symmetric rearrangement of $f_0$ with respect
to $x$. The rearrangement does not 
change the kinetic energy, nor the Casimir functional. 
By \cite[Thm.~3.7]{LL} it can only decrease the potential energy,
more precisely,
\[
-\int\!\!\!\int \frac{f_0(x,v) f_0(y,w)}{|x-y|}dx\,dy
\geq
-\int\!\!\!\int \frac{f_0^*(x,v) f_0^*(y,w)}{|x-y|}dx\,dy
\]
for almost all $v,w\in \R^3$.
But since $f_0$ already minimizes $\H$, $f_0^\ast$ minimizes
$\H$ as well and the potential energy remains unchanged under
the rearrangement which implies that in the above,
equality holds for almost all
$v,w \in \R^3$. By \cite[Thm.~3.9]{LL} this can happen only if
$f_0(x,v)=f_0^\ast(x+T_v,v)$ for some possibly
$v$-dependent shift vector $T_v$. Since both
$f_0$ and $f_0^\ast$ are minimizers they are both of the form
stated in Theorem~\ref{propminim}, 
so $f_0(x,v)=\phi(E_0-\frac{1}{2}|v|^2-U_{f_0}(x))$
and $f_0^\ast(x,v)=\phi(E_0^\ast-\frac{1}{2}|v|^2-U_{f_0^\ast}(x))$.
The explicit form of $E_0$ now implies that $E_0=E_0^\ast$, hence  
$U_{f_0}(x)=U_{f_0^\ast}(x+T_v)$ and the translation $T_v$ is
independent of $v$.
Hence the minimizer $f_0$ is a spatial translation
of $f_0^\ast$ which proves the symmetry assertion.
In passing we note that the symmetry can also be obtained without the 
rearrangement concept, cf.~\cite{R4}.

To prove part (b),
consider first the case where $k<1/2$ i.~e., $n<2$. Then 
$\rho_0 \in L^p \cap L^1$ with $p=1+1/n > 3/2$. The usual 
$L^p$-regularity theory and Sobolev's embedding theorem
implies that 
\[
U_0 \in W^{2,p}_{\rm loc} (\R^3) \subset C(\R^3).
\]
Moreover, for any $R>0$ and $x \in \R^3$,
\beas
- U_0 (x)
&=&
\int_{|x-y| < 1/R} \frac{\rho_0 (y)}{|x-y|}dy
+  \int_{1/R \leq |x-y| < R} \cdots +  \int_{|x-y| \geq R} \cdots\\
&\leq& 
C \|\rho_0\|_p \left( \int_0^{1/R} r^{2-q}dr\right)^{1/q}
+
R \int_{|y| \geq |x| - R} \rho_0(y)\, dy +\frac{M}{R},
\eeas
where $q$ is the conjugate exponent to $p$, so $q<3$.  
This implies that 
\[
U_0 \in C_b(\R^3),\ \lim_{|x| \to \infty} U_0 (x) = 0.
\]
This in turn implies that for $|x|$ sufficiently large,
$E > E_0$; note that the latter quantity is negative by
Theorem~\ref{propminim}. By the same theorem,
$f_0$ and $\rho_0$ have compact support. 

To continue, we note that since $f_0$ depends 
only on the particle energy $E$ via the function $\phi$, 
\be \label{rhorep}
\rho_0 (x) = h_\phi (U_0 (x)),\ x \in \R^3
\ee
where
\be \label{hdef}
h_\phi (u) :=  
4 \pi \sqrt{2} \int_u^\infty \phi(E)\, \sqrt{E-u}\, dE,\
u \in \R;
\ee
note that $h_\phi (u) = 0$ for $u \geq E_0$.
By the general assumptions on $Q$ the function $h_\phi$ is continuously
differentiable. Thus the regularity of $U_0$ implies that
$\rho_0 \in C_c(\R^3)$, this in turn implies that 
$U_0 \in C^1_b(\R^3)$, thus 
$\rho_0 \in C^1_c(\R^3)$, and finally $U_0 \in C^2_b(\R^3)$.

Consider now the case that $k \geq 1/2$. Clearly we are done if we
can but prove that $\rho_0$ is not only in 
$L^p$ with $p=1+1/n$, which is now to small for the argument above,
but in some $L^p$ with $p> 3/2$. To show this, we use
a bootstrap argument, based on (\ref{rhorep}). For this to work,
we need some control on the growth of the function $h_\phi$
which is the reason for the extra assumption on $\phi$. 
Indeed, under that assumption the following estimate holds: 
\[
h_\phi (u) \leq C \left( 1+(E_0 - u)^n \right),\ u \leq E_0.
\]
If we use this estimate on the set where $\rho_0$ is large---this set 
has finite measure---and the integrability of $\rho_0$ on the complement
we find that
\be \label{boot}
\int \rho_0 (x)^p \, dx \leq C + \int (-U_0 (x))^{n p} dx .
\ee
If we would pick the limiting case $n=5$, i.~e., $\rho_0 \in L^{6/5}$,
we would find by Young's inequality
that $U_0 \in L^6$, and bootstrapping this via (\ref{boot}) gives us
$\rho_0 \in L^{6/5}$ back. However, for $n<5$ this works better:
Starting with $p_0=1+1/n$ we apply Young's inequality to find that
$U_0$ lies in  $L^q$ with $q=(1/p_0 - 2/3)^{-1}> 1$,
and substituting this into (\ref{boot}) we conclude that $\rho_0 \in L^{p_1}$
with $p_1=q/n$; note that by assumption $p_0 < 3/2$. If $p_1 >3/2$
we are done. If $p_1 =3/2$ we decrease $p_1$ slightly---note
that $\rho_0 \in L^1$---so that in the next bootstrap step we find
$p_2$ as large as we wish. If $p_1 < 3/2$ we repeat the process.
By induction one sees that
\[
p_k = \frac{3(1+1/n)(n-1)}{n^k (n-5) + 2n + 2} > 1
\]
as long as $p_{k-1} < 3/2$. But since $2\leq n<5$ the denominator 
would eventually become negative so that the process must stop
after finitely many steps.

As to part (c) we first observe that up to some shift $U_0$
as a function of the radial variable $r:=|x|$ solves the equation
\be \label{ef}
\frac{1}{r^2} (r^2 U_0')' = c_k (E_0 - U_0)_+^{k+3/2},\ r>0,
\ee
with some appropriately defined constant $c_k$. Here $'$ denotes the
derivative with respect to $r$. The assertion now follows from the
scaling properties of (\ref{ef}), and we refer to \cite[Thm.~3]{G2}
for the details.
\prfe

\section{Dynamical stability}
\setcounter{equation}{0}

Let $f_0 \in \F_M$ be a minimizer as obtained in Theorem~\ref{exminim}.
To investigate its dynamical stability we note first
that 
\bea  \nonumber
\H (f)- \H (f_0)
&=&
\int\!\!\int \left(\frac{1}{2} |v|^2 + U_0 \right)(f-f_0)  \,dv\,dx 
- \frac{1}{8 \pi} \|\nabla U_f-\nabla U_0\|^2_2\\
&=:&
d(f,f_0) 
- \frac{1}{8 \pi} \|\nabla U_f-\nabla U_0\|^2_2,\ f \in \F_M
\label{d-d} .
\eea
Since $\C(f)=\C(f_0)$, 
\[
d(f,f_0) =
\int\!\!\int
\left[E (f-f_0)- \lambda_0 (Q(f)-Q(f_0))\right]\,\,dv\,dx.
\]
Since $Q$ is convex and the Lagrange multiplier
$\lambda_0$ from Theorem~\ref{propminim} is negative, 
the integrand can be estimated from below by
\[
(E - \lambda_0 Q'(f_0))\,(f-f_0).
\]
According to Theorem~\ref{propminim},
this quantity is zero
on $\supp f_0$, while on
$\R^6 \setminus \supp f_0$ it equals
\[
(E  - \lambda_0 Q'(0))\, f = (E-E_0)\, f \geq 0.
\]
Thus we see that
\[
d(f,f_0) \geq 0,\ f \in \F_M.
\]

We are now ready to state our stability result. Note that
if we shift a minimizer in space we obtain another
minimizer. Moreover, we do in general not know 
whether the minimizers are unique up to spatial shifts.
This fact is reflected in two versions of our stability result.

\begin{theorem} \label{stability}
Let $\M_M \subset \F_M$ denote the set of all minimizers of
$\H$ in $\F_M$.
\begin{itemize}
\item[{\rm (a)}]
For every $\epsilon>0$ there is a $\delta>0$ such that
for any solution $t \mapsto f(t)$ of the Vlasov-Poisson system
with $f(0) \in C^1_c (\R^6)\cap \F_M$,
\[
\inf_{f_0 \in \M_M} 
\left[ 
d(f(0),f_0) + \frac{1}{8\pi} \|\nabla U_{f(0)}-\nabla U_0\|_2^2
\right] < \delta
\]
implies that
\[
\inf_{f_0 \in \M_M}
\left[ 
d( f (t),f_0) + \frac{1}{8\pi} 
\|\nabla U_{ f (t)}-\nabla U_{f_0}\|_2^2 \right]< \epsilon,\ t \geq 0.
\]
\item[{\rm (b)}]
Suppose that $f_0 \in \M_M$ is isolated, i.~e.,
\[
\inf\left\{\| \nabla U_{f_0} - \nabla U_{\tilde f_0} \|_2
\mid
\tilde f_0 \in \M_M\setminus \{T^a f_0 \mid a \in \R^3\}\right\}
>0.
\]
Then for every $\epsilon>0$ there is a $\delta>0$ such that
for any solution $t \mapsto f(t)$ of the Vlasov-Poisson system
with $f(0) \in C^1_c (\R^6) \cap  \F_M$,
\[
d(f(0),f_0) + \frac{1}{8\pi} \|\nabla U_{f(0)}-\nabla U_0\|_2^2
< \delta
\]
implies that for every $t \geq 0$ there exists a shift vector $a \in \R^3$
such that
\[
d(f(t), T^a f_0) + \frac{1}{8\pi} 
\|\nabla U_{f(t)}-\nabla U_{T^a f_0}\|_2^2 < \epsilon.
\]
Here $T^a f (x,v) := f(x+a,v)$ for $a \in \R^3$.
\end{itemize}
\end{theorem}

\noindent
{\bf Remark.}
\begin{itemize}
\item[(a)]
By Theorem~\ref{regminim} (c) the assumption of part (b)
holds for the polytropes.
\item[(b)]
We only showed that $d(f,f_0)\geq 0$ for $f \in \F_M$, but one may 
think of this term as a weighted $L^2$-difference of $f$ and $f_0$.
For example, 
if $Q \in C^2(]0,\infty[)$ with
\[
c_Q := \inf_{0<f\leq f_{\rm max}} Q''(f) > 0
\]
where $f_{\rm max} \geq \|f_0\|_\infty$
as would be the case
for the polytropes $Q(f) = f+ f^{1+1/k}$ with $1\leq k <7/2$,
then by Taylor-expanding $Q$ we find
\[
d(f,f_0) \geq \frac{1}{2} c_Q  \|f-f_0\|_2^2,\ f \in \F_M
\ \mbox{with} \ f \leq f_{\rm max};
\]
observe that the size restriction on $f$ propagates
along solutions of the Vlasov-Poisson system.
\item[(c)]
The restriction $f(0) \in \F_M$ for the perturbed initial data is
acceptable from a physics point of view: A physical perturbation
of a given galaxy, say by the gravitational pull of some outside object,
would result in a perturbed state which is an equimeasurable rearrangement
of the original state, in particular, the value of $\C(f)$ remains unchanged.
\end{itemize}

\noindent
{\bf Proof of Theorem~\ref{stability}.}
Assume the assertion of part (a) were false. 
Then there exist $\epsilon>0$, $t_n>0$, and
$f_n(0) \in C^1_c (\R^6) \cap \F_M$ such that
for all $n \in \N$, 
\be \label{dist0}
\inf_{f_0 \in \M_M} 
\left[ 
d(f_n(0),f_0) + \frac{1}{8\pi} \|\nabla U_{f_n(0)}-\nabla U_0\|_2^2
\right] < \frac{1}{n}
\ee
but
\be \label{disttn}
\inf_{f_0 \in \M_M}
\left[ d(f_n (t_n),f_0) + \frac{1}{8\pi} 
\|\nabla U_{f_n(t_n)}-\nabla U_0\|_2^2 \right] \geq \epsilon.
\ee
By (\ref{dist0}) and (\ref{d-d}),
\[
\lim_{n\to \infty}\H(f_n (0)) = h_M.
\]
Since both $\H$ and $\C$ are conserved along classical solutions as
launched by $f_n (0)$,
\[
\lim_{n\to \infty}\H( f_n (t_n)) =  h_M \ \mbox{and}\ 
f_n (t_n) \in \F_M,\ n \in \N,
\]
i.~e., $(f_n (t_n))$ is a minimizing sequence for $\H$ in $\F_M$.
Up to a subsequence we may therefore assume by Theorem~\ref{exminim}
that there exists a minimizer $f_0 \in \F_M$
and a sequence $(a_n) \subset \R^3$ such that
\be \label{fieldn}
\|\nabla U_{f_n (t_n)}-\nabla U_{T^{a_n} f_0}\|^2_2 \to 0;
\ee
note that for any $f \in \F_M$ and $a \in \R^3$,
\[
\|\nabla U_{T^a f}-\nabla U_{f_0}\|_2
=\|\nabla U_{f}-\nabla U_{T^{-a}f_0}\|_2,
\]
also $d(T^a f,f_0)=d(f,T^{-a}f_0)$.
Since 
$\lim_{n\to \infty}\H(f_n (t_n)) = h_M = \H (T^{a_n}f_0)$
we conclude by (\ref{fieldn}) and (\ref{d-d}) that
\[
d(f_n (t_n),T^{a_n} f_0) \to 0,\ n \to \infty,
\]
and since $T^{a_n} f_0 \in \M_M$ we
arrive at a contradiction to (\ref{disttn}). 
Thus part (a) is established.

Now assume that $f_0$ is an isolated minimizer in $\F_M$, and define
\[
\delta_0 := 
\frac{1}{8 \pi}
\inf\left\{\| \nabla U_{f_0} - \nabla U_{\tilde f_0} \|_2
\mid
\tilde f_0 \in \M_M\setminus \{T^a f_0 \mid a \in \R^3\}\right\} > 0.
\]
Let $\epsilon >0$ arbitrary. In order to find the
corresponding $\delta$ we can without loss of
generality assume that
$\epsilon < \delta_0/4$.
Now choose
$\delta > 0$ according to part (a), without
loss of generality $\delta < \epsilon$, and let 
$f(0) \in  C^1_c (\R^6) \cap \F_M$ be such that
\[
d(f(0),f_0) + \frac{1}{8\pi} \|\nabla U_{f(0)}-\nabla U_0\|_2^2
< \delta .
\] 
The function
\beas
h(t,a)
&:=&
d(f(t), T^a f_0) + \frac{1}{8\pi} 
\|\nabla U_{f(t)}-\nabla U_{T^a f_0}\|_2^2 \\
&=&
\int\!\!\!\int \frac{1}{2} |v|^2 (f(t)-f_0)\, dv\, dx
+ \frac{1}{8 \pi} \|\nabla U_{f(t)}\|_2^2 
+ \frac{3}{8 \pi} \|\nabla U_{f_0}\|_2^2 \\
&&
{} +  \int U_{T^a f_0}\rho_{f(t)} dx
\eeas
is continuous,
and since the interaction term goes to zero as $|a| \to \infty$,
uniformly on compact time intervals, $\inf_{a \in \R^3} h(t,a)$
is also continuous. Now
assume that there exists $t>0$ such that
\[
\inf_{a \in \R^3} h(t,a) \geq \epsilon.
\]
Since at time zero the left
hand side is less then $\epsilon$ there exists some
$t^* > 0$ where
\be \label{fndist}
\inf_{a \in \R^3} h(t^\ast,a) = \epsilon.
\ee
On the other hand, part (a) provides some $f_0^\ast \in \M_M$ such that
\be \label{infdist}
d( f (t^\ast),f_0^\ast) + \frac{1}{8\pi} 
\|\nabla U_{ f (t^\ast)}-\nabla U_{f_0^\ast}\|_2^2  < \epsilon \leq
\frac{\delta_0}{4}.
\ee
By (\ref{fndist}) and (\ref{infdist}) together with
the non-negativity of $d$,
\[
\frac{1}{8\pi} 
\|\nabla U_{f_0}-\nabla U_{f_0^\ast}\|_2^2 \leq \frac{\delta_0}{2},
\]
and by the definition of $\delta_0$ there must exist some
$a^\ast \in \R^3$ such that
$f_0^\ast = T^{a^\ast} f_0$.
But this means that (\ref{fndist}) contradicts (\ref{infdist}),
and the proof of part (b) is complete.
\prfe

\section{The case $k=7/2$; the Plummer sphere}
\setcounter{equation}{0}

In this section we study the so-called
Plummer sphere which corresponds to the 
minimization of $\H$ on the constraint set
\[
\F_M := \left\{f \in L^{9/7}(\R^6) |
f \geq 0,\ \int\!\!\!\int f^{9/7} dv\, dx = M,\ \ekin(f) < \infty \right\}
\]
i.~e., we take $Q(f)=f^{9/7}$ which means $k=7/2$ and $n=5$.
Due to the fact that the scaling transformation
\be \label{s}
(S_\lambda f)(x,v)=\lambda^{-7} f(\lambda^{-4} x, \lambda v),
\ee
leaves each term in both ${\cal H}$ and $\C$ invariant
this case poses additional difficulties. 

As we noted at the end of Section~2 the assertion of Lemma~\ref{lower}
remains valid so that there exists a minimizing sequence $(f_i)$
of $\H$ in $\F_M$.
We shall follow the steps in Section~3 to conclude the
existence of a minimizer. 
The key step is to find the analogue of
Lemma~\ref{concentrate} in the presence of the 
scaling (\ref{s}) and to deal with compactness issues in the
limiting case $(\rho_i) \subset L^{6/5}(\R^3)$, cf.\ Lemma~\ref{compact}. 
For a function $h=h(x,v) \geq 0$ and a cut-off
parameter $N>0$ we define
\[
h_{|N} (x,v) :=
\left\{
\begin{array}{cl}
h(x,v) &\ \mbox{if}\ 1/N \le h(x,v) \le N,\\
0      & \ \mbox{otherwise}
\end{array} \right.
\]

\begin{lemma} \label{concentratepl}
Let $(f_i) \subset \F_M$ be minimizing.
Then there exists a sequence $(\lambda_i)\subset ]0,\infty[$ such that
up to a subsequence the following holds
for $S_i f_i := S_{\lambda_i} f_i$: 
\begin{itemize}
\item[{\rm (a)}]
For any $\epsilon >0$ there exists some $N>0$ such that
for all sufficiently large $i \in \N$,
\[
\|S_i f_i - (S_i f_i)_{|N}\|_{9/7} < \epsilon.
\]
\item[{\rm (b)}]
There exist a sequence
$ (a_i) \subset \R^3$ and $\epsilon_0>0,\ R_0>0$ such that 
for all sufficiently large $i \in \N$,
\[
\int_{a_i+B_{R_0}} (S_i f_i)^{9/7}\, dv\,dx \geq \epsilon_0.
\]
\item[{\rm (c)}]
Let $g_i(x,v):= T_i S_i f_i (x,v) := 
\lambda_i^{-7} f_i(\lambda _i^{-4}x+a_i, \lambda_i v)$,
$g_i \rightharpoonup g_0$ weakly in $L^{9/7}(\R^6)$
and $\rho_i := \rho_{g_i} \rightharpoonup \rho_0$
weakly in $L^{6/5}(\R^3)$. 
Then for any $R>0$ and up to a subsequence,
$\nabla U_{{\bf 1}_{B_R} \rho_i} \to \nabla U_{{\bf 1}_{B_R} \rho_0}$
strongly in $L^2(\R^3)$.
\end{itemize}
\end{lemma}

\noindent
{\bf Proof}. 
To prove part (a) we wish to employ the results
in \cite{G2}, so we consider $(f_i^*)$, the sequence of
spherically symmetric rearrangements with
respect to $x$, which is again minimizing and in $\F_M$. 
By \cite[Thm.~2]{G2} there exists a symmetric 
minimizer $g$ such that $S_i f_i^* \rightharpoonup g$ weakly 
in $L^{9/7}(\R^6)$. Since $\|S_i f_i^\ast\|_{9/7} = \|f_i^\ast\|_{9/7}
= M = \|g\|_{9/7}$ it follows that
$S_i f_i^* \to g$ strongly in $L^{9/7}(\R^6)$. Note that
since $S_i f_i^\ast$ is spherically symmetric and decreasing in $x$
and equi-measurable with $S_i f_i$ we have $S_i f_i^\ast = (S_i f_i)^\ast$.
Abusing the notation we abbreviate $g_i:=S_i f_i$ in the proofs
of parts (a) and (b).

For $\epsilon > 0$ we choose $N>0$ such that 
$\|g\|_{L^{9/7}(A_N)} < \epsilon/2$
where
$A_N =\{g \leq  1/N \vee g \geq N\}$.
Let $A_{i,N}=\{g_i^* < 1/N \vee g_i^\ast >N\}$. Then
for $i$ sufficiently large, 
\beas
\|g_i^* - {g_i^*}_{|N}\|_{9/7}
&=&
\| g_i^* \|_{L^{9/7}(A_{i,N})}
\le
\| g_i^* - g \|_{L^{9/7}(A_{i,N})}
+ \| g \|_{L^{9/7}(A_{i,N})}\\
&\le&
\epsilon/2 + \| g \|_{L^{9/7}(A_{i,N})}.
\eeas
Up to a subsequence,
$g_i^* \to g$ pointwise a.~e., and 
thus $\limsup_{i\to\infty}{\bf 1}_{A_{i,N}} \leq {\bf 1}_{A_N}$ a.~e.. 
Therefore, by Fatou's lemma we get
\[
\limsup_{i\to\infty} \| g \|_{L^{9/7}(A_{i,N})} \leq  
\| g \|_{L^{9/7}(A_N)} < \epsilon/2.
\]
Hence up to a subsequence and for $i$ sufficiently large,
\be \label{N1}
\|g_i^* - {g_i^*}_{|N}\|_{9/7} < \epsilon.
\ee
Now by the equi-measurability of the 
rearrangements, (\ref{N1}) implies part (a) of the lemma. In fact, 
for any function $h\geq 0$ and $p \geq 1$,
\beas
\int (h-h_{|N})^p
&=&
p\int_0^\infty s^{p-1} \mu 
\{h {\bf 1}_{\{h<1/N \vee  h>N\}} > s \}\,ds\\
&=&
p\int_0^\infty s^{p-1} 
\Bigl[\mu \{s<h<1/N\} + \mu \{h>\max [N,s]\}\Bigr] \,ds\\
&=&
p\int_0^\infty s^{p-1} 
\Bigl[\mu \{s< h^* <1/N\} + \mu \{h^* >\max [N,s]\}\Bigr] \,ds\\
&=&
\int (h^*-{h^*}_{|N})^p.
\eeas
Taking $h=g_i=S_i f_i$ and recalling that $S_i f_i^\ast = (S_i f_i)^\ast$
proves part (a) of the lemma.

To prove part (b) we split for $N>0$,
$g_i = {g_i}_{|N} + (g_i -{g_i}_{|N})$.
With the bounded sequence $({g_i}_{|N})$ instead of $(f_i)$ we
split as in (\ref{i}) of Lemma~\ref{concentrate}. Since
$({g_i}_{|N})$ is bounded in $L^1 \cap L^\infty$ the induced
spatial densities are bounded in, say, $L^{5/4}$
which we use to bound $I_3$; $I_1$ and $I_2$ are treated as before.
Thus instead of (\ref{ii}) we obtain
\beas
&&
\left[\sup_{y\in \R^3}\int_{y+B_R}
\int (g_i)^{9/7}\,dv\,dx \right]^{7/12} \geq
\left[\sup_{y\in \R^3}\int_{y+B_R}
\int ({g_i}_{|N})^{9/7}\,dv\,dx \right]^{7/12}\\
&&
\hspace*{3.5cm}
\geq
C\, R^{-3/2}
\left[ - h_M /2 - C R^{-1} - C R^{-1/5} - 
\|\nabla U_{g_i-{g_i}_{|N}}\|_2^2 \right] .
\eeas 
Because of part (a) and Lemma~\ref{rhoest}
we can choose $N>0$ such that $\|\nabla U_{g_i-{g_i}_{|N}}\|_2<-h_M/4$. 
Then we choose $R>0$ such that the bracket is
positive, and the proof of part (b) is complete. 

As to part (c) we first show that for any $R>0$ the set 
$\{\nabla U_{{\bf 1}_{B_R} \rho_i} | i \in \N \}$ is relatively compact
in $L^2(\R^3)$. To see this note that we can for any 
$\epsilon>0$ choose $N>0$ and $R' > R$ such that 
for $i$ sufficiently large,
\[
\|{\bf 1}_{B_{R'}}\nabla U_{{\bf 1}_{B_R}{g_i}_{|N}} - 
\nabla U_{{\bf 1}_{B_R} g_i}\|_2 < \epsilon;
\]
cf.\ part (a) together with Lemma~\ref{rhoest}
and the splitting used in the proof of Lemma~\ref{compact}.
For $R'$ and $N$ fixed the set 
$\{{\bf 1}_{B_{R'}}\nabla U_{{\bf 1}_{B_R}{g_i}_{|N}} | i \in \N \}$ 
is relatively compact in $L^2(\R^3)$ since $\{{g_i}_{|N}\}$ is bounded
in $L^1\cap L^\infty (\R^6)$ 
and thus the set of induced densities is bounded, say,
in $L^{5/4}(\R^3)$. This implies the relative compactness stated above.
Next we observe that ${\bf 1}_{B_R}\rho_i \rightharpoonup 
{\bf 1}_{B_R}\rho_0$ weakly in $L^{6/5}(\R^3)$ which implies that
that $\nabla U_{{\bf 1}_{B_R}\rho_i} \rightharpoonup 
\nabla U_{{\bf 1}_{B_R}\rho_0}$
weakly in $L^2(\R^3)$. Together,
this yields the assertion of part (c).  
\prfe

We are now ready to prove the analogue of Theorem~\ref{exminim}
for the limiting case $k=7/2$:

\begin{theorem} \label{exminimp}
Let $M>0$.
Let $(f_i) \subset \F_M$ be a minimizing sequence of 
$\H$. Then there is a minimizer $f_0\in \F_M$, a subsequence (still 
denoted by 
$(f_{i})$), and a sequence of translations and scalings: 
\[
 T_i S_i f_i (x,v) := \lambda_i^{-7} f_i(\lambda _i^{-4}x+a_i, \lambda_i v)
\] 
with $(a_i) \subset \R^3$ and $\lambda_i > 0$ such that 
\[
\H (f_0) = \inf_{\F_M} \H = h_M
\] 
and $ T_i S_i f_{i} \rightharpoonup f_0$ weakly in 
$L^{9/7} (\R^6)$.
For the induced potentials we have
$\nabla U_{T_i S_if_i} \to \nabla U_0$ strongly in $L^2 (\R^3)$.
\end{theorem}

\noindent
{\bf Proof}. We choose $(\lambda_i)$, $(a_i)$, $\epsilon_0>0$,
and $R_0>0$ according to Lemma~\ref{concentratepl} and define 
$g_i := T_i S_i f_i$, which is again a minimizing sequence in $\F_M$.
This sequence is bounded in $L^{9/7}(\R^6)$, and by 
Lemma~\ref{rhoest} (b) the induced spatial
densities $\rho_i$ are bounded in $L^{6/5}(\R^3)$ so that up to
a subsequence,
$g_i \rightharpoonup f_0$ weakly in $L^{9/7}(\R^6)$, and
$\rho_i \rightharpoonup \rho_0$ weakly in $L^{6/5}(\R^3)$. 
We now proceed as in the
proof of Theorem~\ref{exminim} with $(g_i)$ instead
of $(f_i)$, the only difference being
that instead of Lemma~\ref{concentrate} and \ref{compact}
we use the corresponding parts of Lemma~\ref{concentratepl}
in order to prove the strong converge of the fields
(\ref{strong}). \prfe

Next we show the analogues of the assertions of Theorems~\ref{propminim} and
\ref{regminim}:

\begin{theorem}
\label{propminimp}
Let $f_0 \in \F_M$ be a minimizer of $\H$.
Then
\[
f_0 (x,v) = \left\{
\begin{array}{ccl}
\left(E/\lambda_0 \right)^{7/2}
&,& E < 0,\\
0 &,& E \geq 0,
\end{array}
\right.
\]
with 
some constant $\lambda_0<0$, in particular, $f_0$ is a steady state
of the Vlasov-Poisson system.  Moreover, $f_0$ is spherically
symmetric with respect to some point in $x$-space, and  
up to scalings and translations in $x$,
\[
U_0 (r) = -c_0\,(1 + r^2)^{-1/2},\ 
\rho_0 (r) = \frac{3 c_0}{4 \pi} (1 + r^2)^{-5/2},\ r \geq 0,
\]
where the positive constant $c_0$ depends on $\lambda_0$.
\end{theorem}

\noindent
{\bf Proof}. The identity for $f_0$ follows exactly
as in the proof of Theorem~\ref{propminim}.
The spherical symmetry follows as in the proof of 
Theorem~\ref{regminim}. 
By monotonicity
$\lim_{r \to \infty} U_0 (r) \in ]-\infty, 0]$
exists, and since $U_0 \in L^6(\R^3)$ this limit must be zero
so that $U_0$ is a solution of the corresponding
Emden-Fowler equation (\ref{ef}) with $k=7/2$ and $E_0=0$.
The uniqueness up to scalings follows as in 
\cite[Thm.~3]{G2}, and the explicit formulas can be checked
by direct computation. \prfe

Finally, we state the stability theorem for the limiting case
$k=7/2$:

\begin{theorem}
\label{stabp}
Let $f_0\in \F_M$ be a minimizer of $\H$. Then for every
$\epsilon>0$ there is a $\delta>0$ such that for any solution 
$t \mapsto f(t)$ of the Vlasov-Poisson system 
with $f(0)\in C^1_c \cap \F_M$, 
\[
d(f(0),f_0)+{1\over{8\pi}}
\|\nabla U_{f(0)}-\nabla
 U_{f_0}\|^2_{L^2} < \delta
\]
implies that for every $t \geq 0$ there exits a shift vector
$a \in \R^3$ and a scaling parameter $\lambda > 0$ such that 
\[
d(f(t),T^aS_\lambda f_0)+{1\over{8\pi}}
\|\nabla U_{f(t)}-\nabla U_{T^a S_\lambda f_0}\|^2_2
<\epsilon,\ t \geq 0.
\]
\end{theorem}

The only difference to the proof of Theorem~\ref{stability}
is that one now has to take into account not only the spatial
shifts, but also the scaling transformations which arise
in Theorem~\ref{exminimp}, and this is straightforward. 
The condition 
$f(0)\in {\cal F}_M$ can also be relaxed by a scaling transformation as 
in \cite[Thm.~4]{G2}.
 
{\bf Note added in proof.}
The stability of isotropic steady states is also addressed in\\
Wan, Y.-H.: Nonlinear stability of spherical systems in galactic dynamics.
Preprint, 2000

\end{document}